\documentclass[a4paper]{spie}  
\usepackage[top=1in, bottom=4.94cm, left=1.925cm, right=1.925cm]{geometry}
\usepackage[]{graphicx}
\usepackage[caption=false,font=normalsize,labelfon
t=sf,textfont=sf]{subfig}
\usepackage{amsmath}

\title{A novel TOF PET reconstruction method from limited-view data based on TV-minimization}

\author{Deng Zhenzhou\supit{a}
\skiplinehalf
\supit{a}Nanchang University, 999 Xuefu Avenue, Nanchang, China
}

\authorinfo{
Deng Zhenzhou: E-mail: zzdeng@ncu.edu.cn}

\begin{document}\small
  \maketitle
  \linespread{1.0}
\begin{abstract}
PET detectors provide the information of the position, the energy and the timing about the decay events on the LOR. Traditional PET image reconstruction has not taken the timing information into account, only used the timing information for coincidence judgments. The high timing resolution PET detectors provide very precise TOF information, then TOF image reconstruction method which utilizes the timing information of PET detectors is so crucial for the TOF PET system. We take advantage of timing information provided by a pair of TOF PET detectors, and then reconstruct the activity distribution from the limited-view projection data. Since the image reconstruction from the limited-view data is an under-determined problem mathematically, conventional algorithms cannot achieve the exact reconstruction of the limited-view problem. In this work, we propose a half-analytic and half-iterative method named DF/LBM (Direct Fourier and Logarithmic Barrier Method) to solve the limited-view problem with the TOF information. The least square solution is obtained via DFM (Direct Fourier Method), and then a convex optimization method named logarithmic barrier method is employed to correct the least square solution. The substance of the convex optimization is defining the components in the null-space to satisfy some prior information (TV minimization), while the least square solution has no components in the null-space. Applying this method to Zubal phantom, the artifact generated by the least square solution is eliminated, and PSNR is 30.6112, 34.8751 and 38.8998 when the timing resolution is unused, 200ps and 100ps respectively and 16 views from $-45$ to $45$ are adopted.We also demonstrate that in certain situations, the new reconstruction algorithms can generate images having improved statistics by recruiting suitable subsets of the TOF-PET data to minimize the use of deteriorating measurements in reconstruction. Potential implications of the new reconstruction approach to PET imaging are discussed.
\end{abstract}

\keywords{Limited-View image reconstruction, Dual-Head, TV minimization, Timing resolution}

\section{INTRODUCTION}
\label{sec:intro}  
PET(positron emission tomography) detectors provide the information of position, energy and timing. Since the timing resolution of traditional PET system is too low to add extra information in image reconstruction, the timing information is only used for coincidence judgments. Now the high timing resolution PET detectors$[1]$ provide precise TOF (time of fly) information, which is fit for TOF PET image reconstruction more than coincidence judgments. Timing information can locate the position of the decaying events in some region. This work proposes a novel image reconstruction method to utilize this ability of locating events on the LOR (line of response). In the system of the Dual-Head PET, the projection views of the data are incomplete. The limited-view data always cause missing-data artifact by conventional algorithms. Consequently, despite dual-head PET has advantages in flexibility, costing and reliability, the applications of dual-head are limited.\\
In this work, we propose a half-analytic and half-iterative method named DF/LBM (Direct Fourier and Logarithmic Barrier Method) to solve the limited-view problem. The incomplete character of limited-view is confirmed for the TOF projection data. Additionally, the least square solution is obtained when projection data are expressed in the Fourier domain. Then the least square solution is corrected by a convex optimization method$[2]$, which is called logarithmic barrier method. The objective function is defined to minimize the TV (total variation)$[3-5]$ of the image function. Applying this method to Zubal phantom, the artifact generated by the least square solution is eliminated, and PSNR is 30.6112, 34.8751 and 38.8998 when the timing resolution is unused, 200ps and 100ps respectively.
\section{Method}
\subsection{The multi-pixel-driven model for TOF PET System}
For 2-D activity image, TOF PET has the 2-D data at each view. Different from the ray tracing mode, the pixel-driven model for TOF PET System is proposed for each view data as follow. We denote the data $p_{\theta}(s,t)$, where $\theta$ is the projection view. $(s,t)$ is the coordinate system which  $t$-axis perpendicular to $\vec{\theta}$ and $t$-axis parallel to $\vec{\theta}$. The Dual-head system can be equivalently described as a FOV (Field Of View) with circle shape between two lines [Fig.1(a)]. LOR connects two points in the two lines respectively. Let $O$ be the origin in the FOV coordinates $(x,y)$ and the geometric center of the Dual-Head system. Therefore any view can be denoted as $\theta(k)$, and any LOR can be denoted as $(l,k)$, where $l$ is the length between $O$ and the midpoint of the LOR, and $k$ is the horizontal distance of two endpoints of the LOR. Hence, the projection data collected by this system can be denoted as $p_{\theta}(s,t)$. Let $r$ be the radius of the FOV, and $w$ be the length of the two panels. To avoid the truncated condition, the maximum $k$ is
\begin{equation}
\label{maxk}
k_{max}=d\times\tan[\arctan(\frac{w}{d})-\arctan(\frac{2r}{\sqrt{w^{2}+d^{2}-4r^{2}}})].
\end{equation}
When $r<<w$, $k_{max}\approx w$, the maximum half allowable angle is $\arctan(w/d)$. Let $f(x,y)$ denote the image function with zero value outside the FOV. The projection $p_{\theta}(s,t)$ can be written as
\begin{equation}
\label{tof projection}
p_{\theta}(s,t)=\int f(x,y)ae^{((s'-s)/\sigma)^{2}}ds'
=\int\int  f(x,y)\psi_{\theta}(x'-x,y'-y)dx'dy'.
\end{equation}
where $a$ is the normalization constant, so that the total probability equal to $1$, $\sigma$ is the uncertainty of the location on the LOR defined by timing information. To express Eq.(2) discretely, we build the linear equation
\begin{equation}
\label{tof projection equation}
p_{\theta}=f*\psi_{\theta}.
\end{equation}
\begin{figure}[!t]
\centerline{\subfloat[] {\includegraphics[width=2.5in]{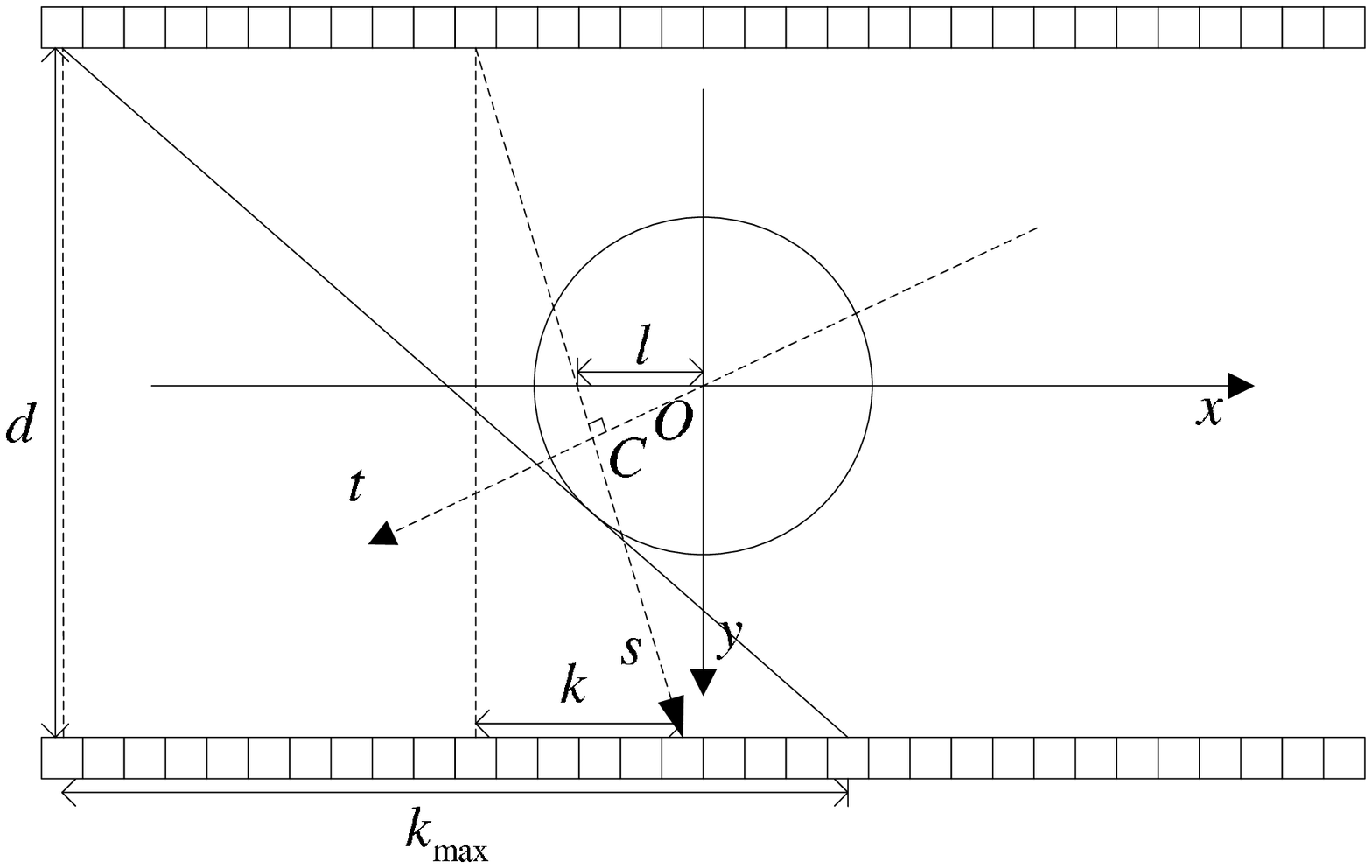}%
\label{LORs in 3 different views}}
\hfil
\subfloat[]{\includegraphics[width=2.5in]{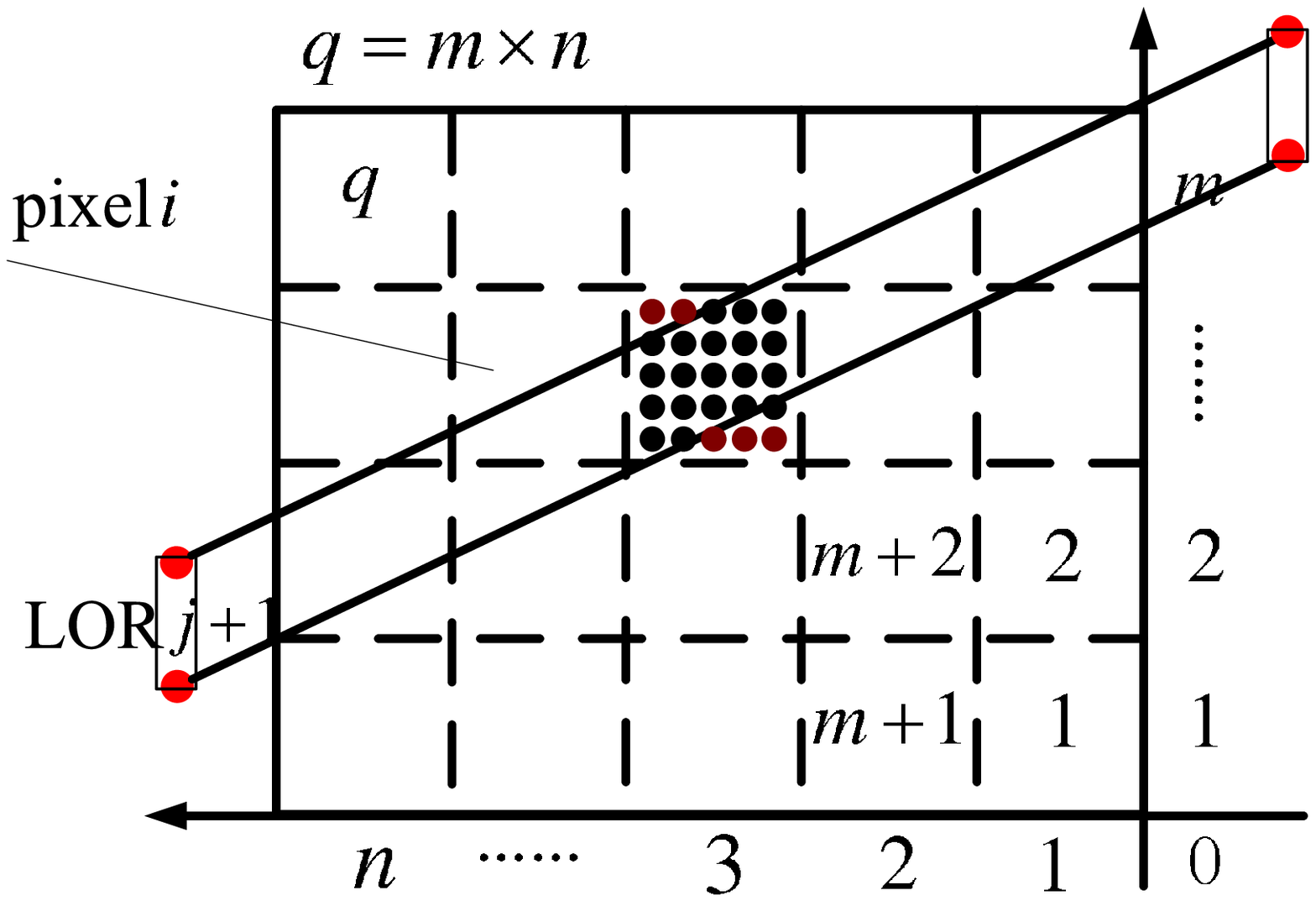}%
\label{LORs in Fourier Expression}}}
\caption{(a) is the Dual-Head TOF PET scanner. (b) shows the sub-pixels divide from one pixel}
\label{fig_sim}
\end{figure}
To count the PSF (pixel spread function) more precisely, we divide the pixel into many tiny pixels [Fig.1(b)]. Then compute the sum of the contributions of the tiny pixels which belong to the pixel. The PSF Result for timing resolution 60ps ($1\sigma$) on view of 0, 45, 60, 135 degree is compute as Fig.2 (a)-(d). The pixel size is 3mm. Fig.2 (f)-(i) is the corresponding projection data.
\begin{figure}
\begin{center}
\begin{tabular}{lllll}
&
\includegraphics[height=3cm]{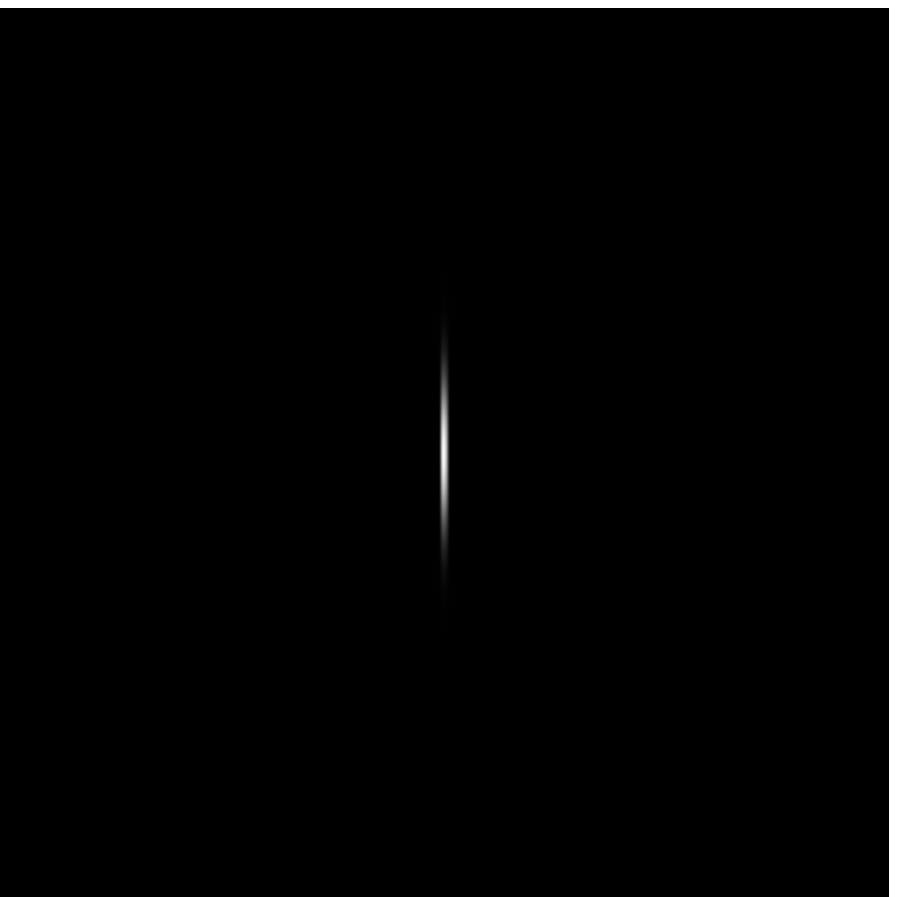}&
\includegraphics[height=3cm]{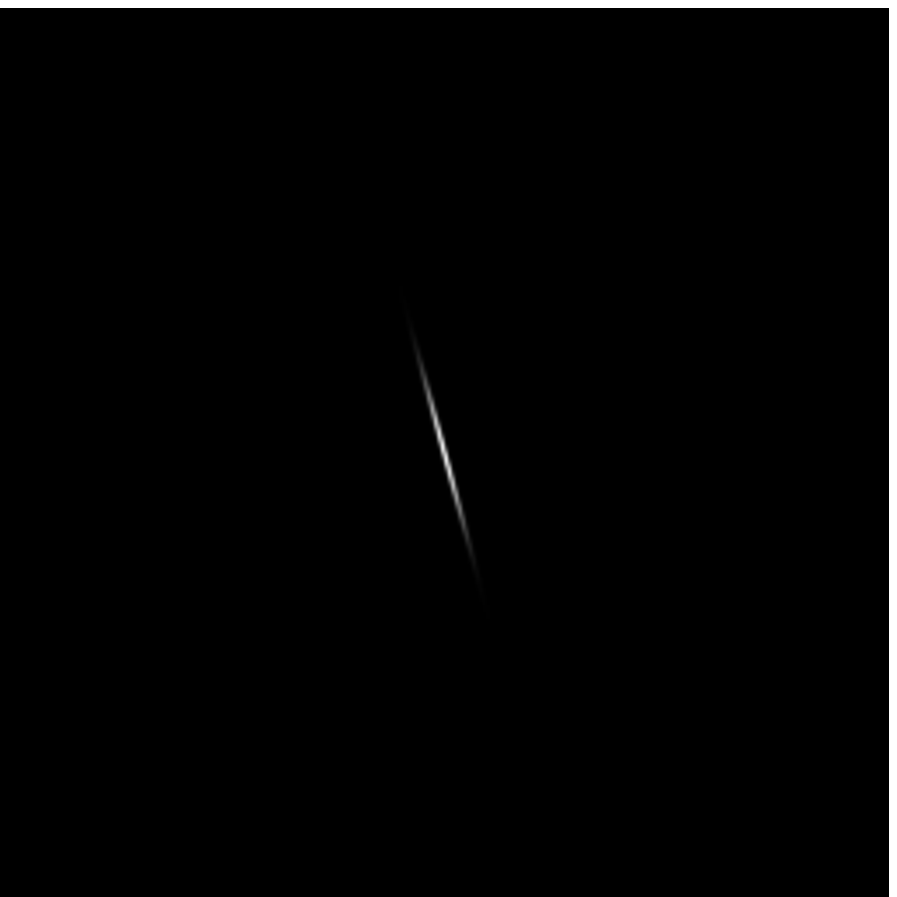}&
\includegraphics[height=3cm]{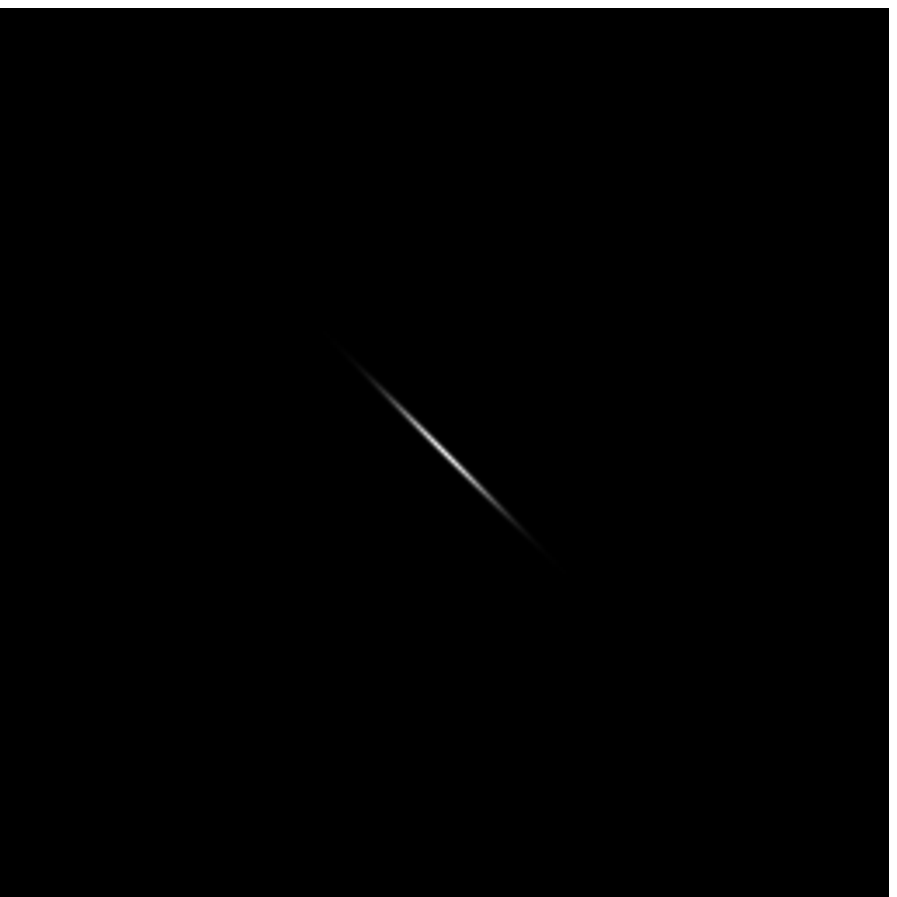}&
\includegraphics[height=3cm]{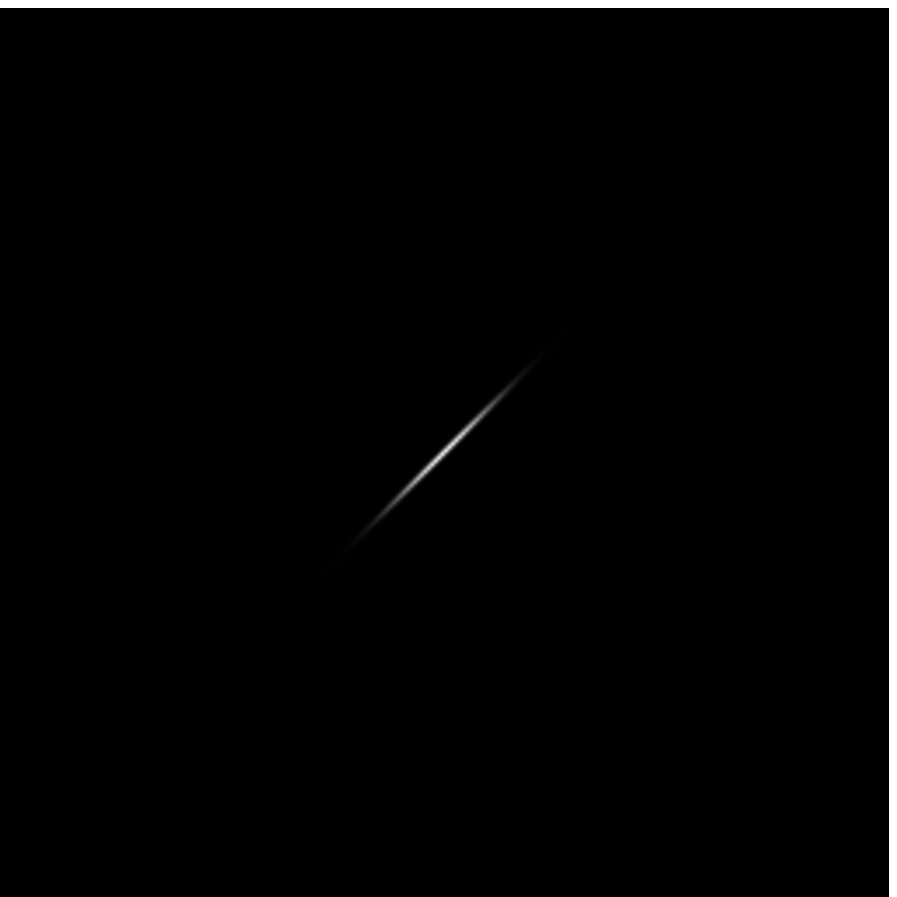}\\
\includegraphics[height=3cm]{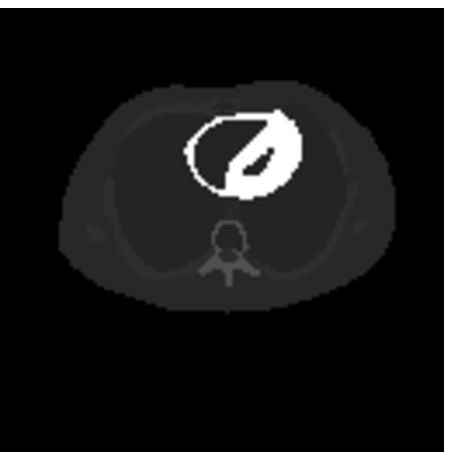}&
\includegraphics[height=3cm]{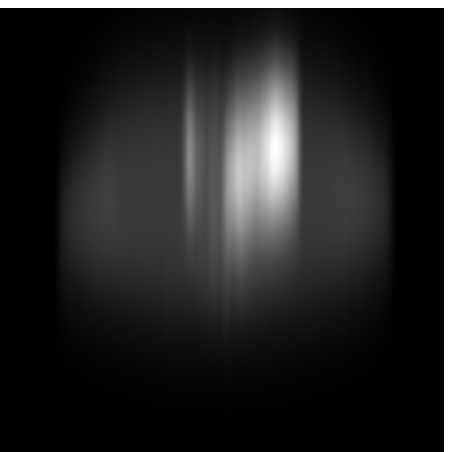}&
\includegraphics[height=3cm]{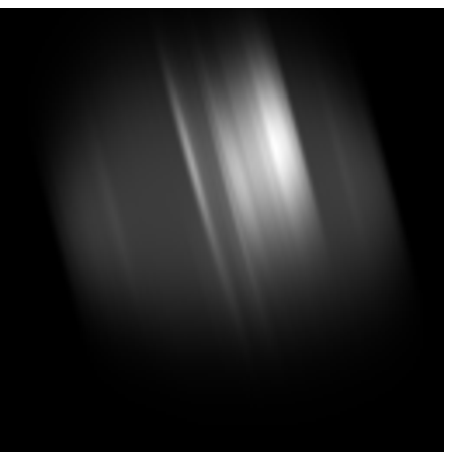}&
\includegraphics[height=3cm]{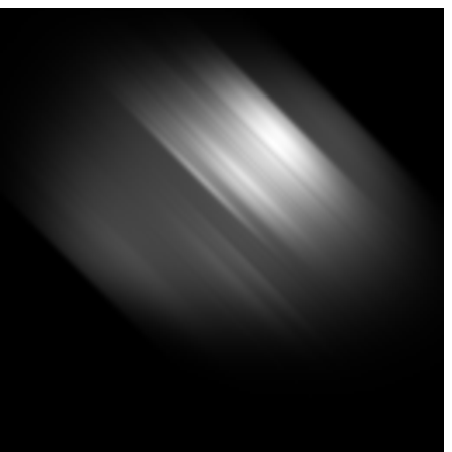}&
\includegraphics[height=3cm]{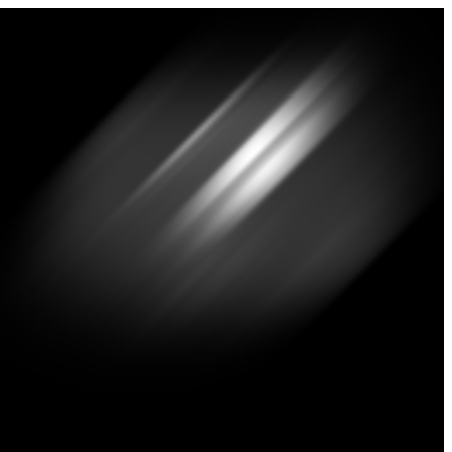}

\end{tabular}
\end{center}
\caption[example]
   { \label{fig:example}
The up row of the pictures are (a)-(d), these pictures show the PSF on the view 0, 15, 45, 135 degree. The PSF are compute by 16 tiny sub-pixels.The down row of the pictures are (e)-(i). (e) is the true distribution of the activity,and (f)-(i) are the TOF project data corresponding (a)-(d). }
   \end{figure}
\subsection{The least square solution of the system matrix}
Components in null-space correspond the non-sampled region in the Fourier domain of the image function [Fig.3(b)]. The point just like $Z$ is given any value, the image function is still satisfied the projection data. Employ DFM (Direct Fourier Method), and the least square solution is obtained by assuming the components in null-space is zero. Adopting the timing information, sampled lines will be wider than the lines in [Fig.3(b)].
\begin{figure}[!t]
\centerline{\subfloat[] {\includegraphics[width=2in,height=3.6cm]{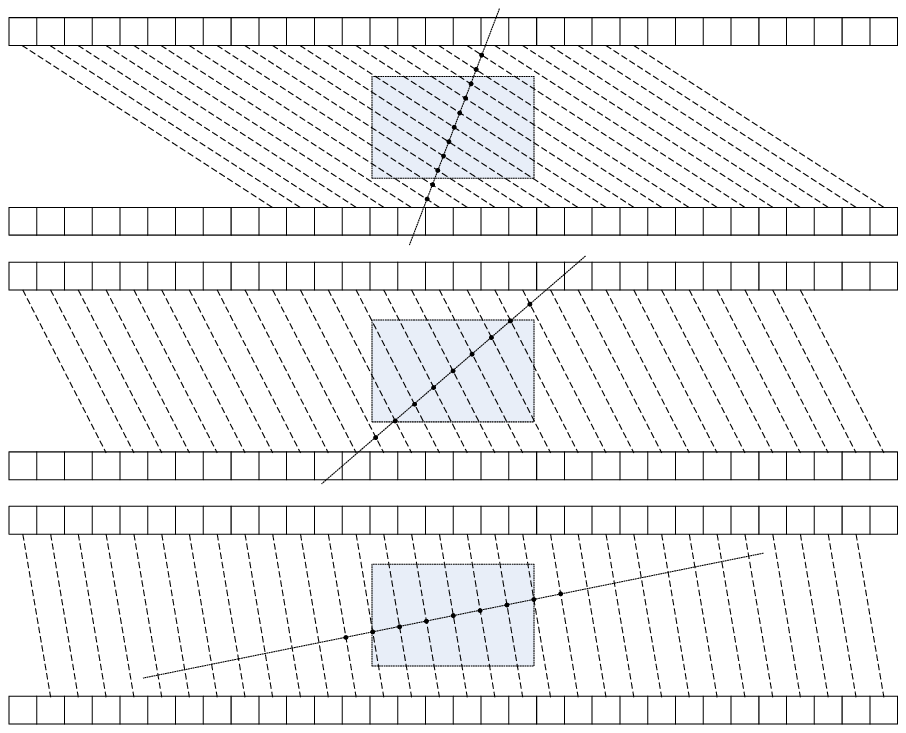}%
\label{LORs in 3 different views}}
\hfil
\subfloat[]{\includegraphics[width=1.4in,height=3.6cm]{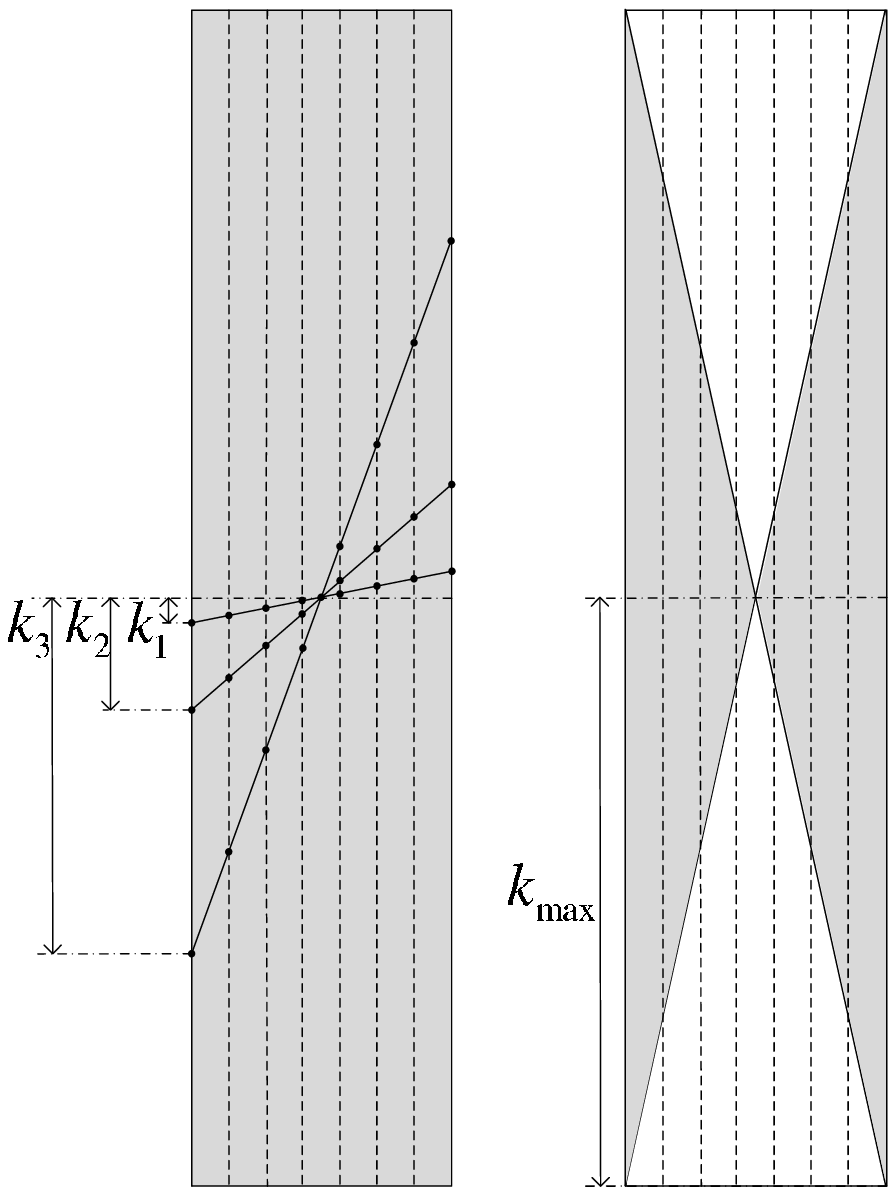}%
\label{LORs in Fourier Expression}}}
\caption{The Fourier Expression of Data in different views}
\label{fig_sim}
\end{figure}
\subsection{Define the components in the null-space}
To choose the components in null-space, the convex optimization method is employed for reconstructing the image function by correcting the least square solution. The TV minimization problem can be written as
\begin{equation}
\label{maxk}
\underset{f}{\min} \underset{i,j}{\sum}\parallel\nabla f(x,y)\parallel
\textsl{~~s.t.~~}
\parallel hf-p\parallel\leq\epsilon,
\end{equation}
where $\nabla$ denotes the gradient operator, and the parameter $\epsilon$ depends on the noise level of the data. The objective function of Eq.(4) is a quadratic polynomial. We rewrite Eq.(4) as
\begin{equation}
\label{maxk}
\underset{t,f}{\min} \underset{x,y}{\sum} t(x,y)
\textsl{~~s.t.~~}\{
\begin{array}{c}
  \parallel hf-p\parallel\leq\epsilon, \\
  \parallel\nabla f(x,y)\parallel^{2}-(t(x,y))^{2}\leq 0.
\end{array}
\end{equation}
Eq.(5) can be abbreviated to be
\begin{equation}
\label{maxk}
\underset{w}{\min} c^{T}w
\textsl{~~s.t.~~}
\varphi(w)\leq 0,
\end{equation}
where $w$ is the connective vector contained $f$ and $t$. This  problem can be solved as unconstrained optimization as follow.
\begin{equation}
\label{maxk}
\underset{w}{\min}~~c^{T}w + \underset{i}{\sum} I_{-}[\varphi_{i}(w)],~~I_{-}(u)=\left\{
                                                                                  \begin{array}{ll}
                                                                                    0, & \hbox{u$\leq$0;} \\
                                                                                    \infty, & \hbox{u$>$0.}
                                                                                  \end{array}
                                                                                \right.
\end{equation}
To make the second term of Eq.(7) derivable, the indicator function $I_{-}$ is approximated by a smooth function $\hat{I}_{-}$, where
\begin{equation}
\label{maxk}
\hat{I}_{-} = -(1/r)log(-u),~~r>0.
\end{equation}
We solve a serious of unconstrained optimizations of the form Eq.(7) as the parameter $r$ grows. Each iteration starts at the solution of the problem for the previous parameter of $r$.
\section{Results}
We apply the algorithm to Zubal phantom. The allowable angle is $-45$ to $45$ degree. Number of the views is $16$. The pictures in the top row of [Fig.4] are DFM's results without components in null-space. The ones in the bottom row are results with components in null-space via minimize TV, which are followed by evaluation of the results [TABLE I].the timing resolution is unused, 200ps and 100ps respectively.

   \begin{figure}[!t]
   \begin{center}
   \begin{tabular}{lll}
   \includegraphics[height=3cm]{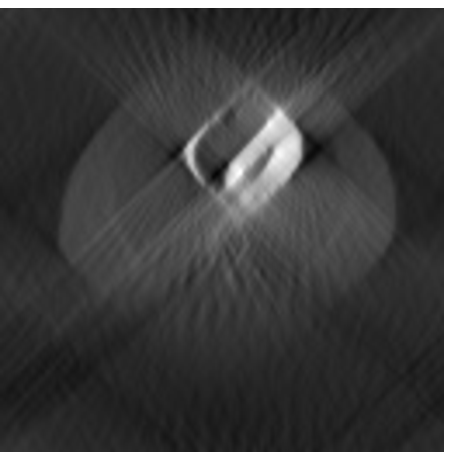}&
   \includegraphics[height=3cm]{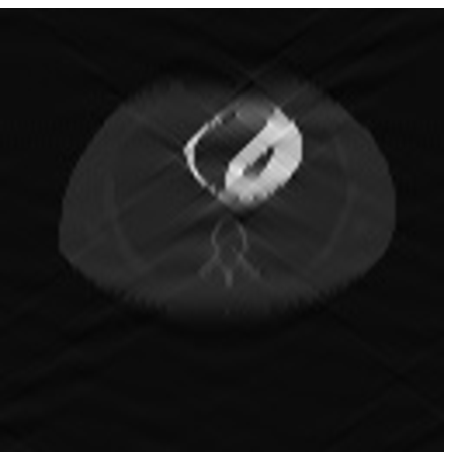}&
   \includegraphics[height=3cm]{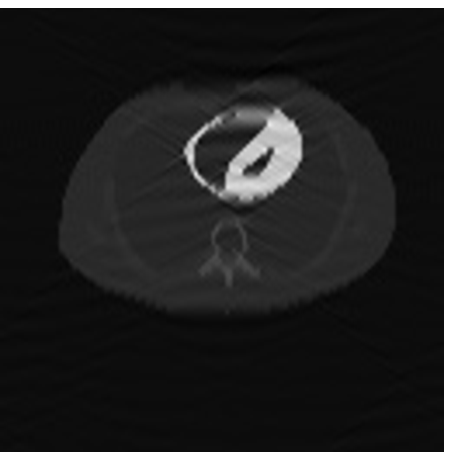}\\
   \includegraphics[height=3cm]{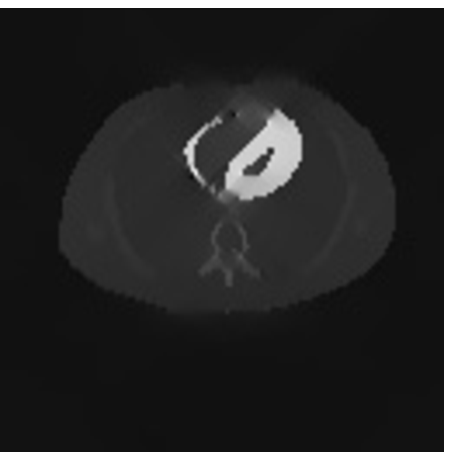}&
   \includegraphics[height=3cm]{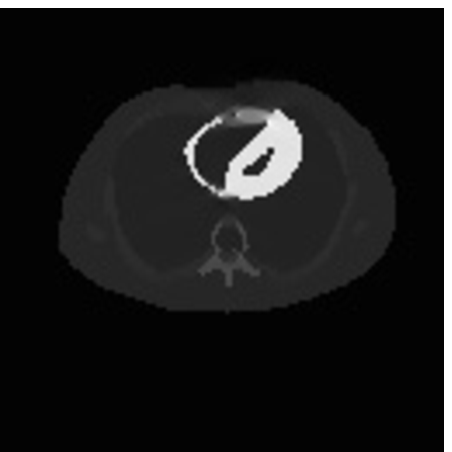}&
   \includegraphics[height=3cm]{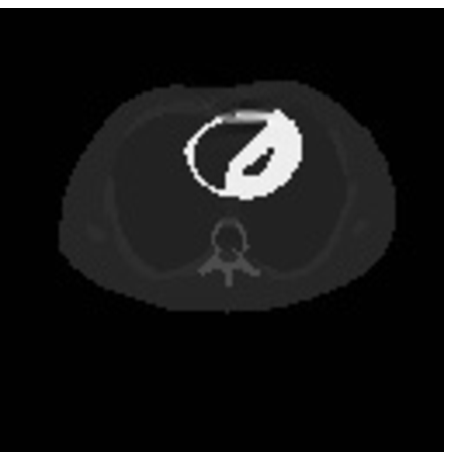}

   \end{tabular}
   \end{center}
   \caption[example]
   { \label{fig:example}
The up row of the pictures are (a)-(c), these pictures show the results of image reconstruction without TV minimization. The down row of the pictures are (d)-(f),these pictures show the results of image reconstruction with TV minimization. (a),(d) do not use the TOF information,(b),(e) use 200ps($1\sigma$) timing resolution and (c),(f) use the 100ps($1\sigma$) timing resolution. }
   \end{figure}
\begin{table}[!th]
\caption{Evaluation of reconstruction results}
\renewcommand{\arraystretch}{1.3}

\label{table_example}
\centering
\begin{tabular}{c c c c c c}
\hline
Timing Resolution & using TV minimization & PSNR  & MSE & MAXERR & L2RAT\\
\hline
unused  & No  & 24.2616 &243.7353 & 180.9633 & 0.7908 \\
unused  & Yes & 30.6112 &56.4881  & 164.1134 & 0.9275 \\
200ps   & No  & 29.3131 &76.1678  & 153.0976 & 0.9346 \\
200ps   & Yes & 34.8751 &21.1629  & 125.8067 & 0.9723 \\
100ps   & No  & 31.7949 &43.0125  & 129.8402 & 0.9631 \\
100ps   & Yes & 38.8998 &8.3772   & 96.3903  & 0.9883 \\
\hline
\end{tabular}
\end{table}
\section{Conclusion}
We proposed a novel algorithm of TOF PET image reconstruction based on TV minimization to solve the problem of dual-head detection geometry. The artifact generated by the least square solution is eliminated, and PSNR is 30.6112, 34.8751 and 38.8998 when the timing resolution is unused, 200ps and 100ps respectively.The simulation results show superiority of this algorithm on solving under-determined problems.
\section{Novelty of this work to be presented}
We proposed a novel algorithm of TOF PET image reconstruction. Limited-View is an under-determined problem mathematically, the optimization method can choose the null-space components to satisfy some object function. Timing information just right enlarges the sample region in fourier domain, and then make the reconstruction more precise.\\


\begin{thebibliography}{2}
\bibitem
1K. Shibuya, H. Saito, M.Koshimizu, K. Asai,
\lq\lq Outstanding Timing Resolution of Pure CsBr Scintillators for Coincidence Measurements of Positron Annihilation Radiation,"
\emph{Applied physics express},
vol.3, pp.086401, 2010.
\bibitem
1S. Boyd, L. Vandenberghe,
\lq\lq Convex Optimization,"
\emph{Cambridge University Press},
2004.
\bibitem
1E. Y. Sidky, C. M. Kao, X. Pan,
\lq\lq Accurate image reconstruction from few-views and limited-angle data in divergent-beam CT,"
\emph{Journal of X-Ray Science and Technology},
vol.14, pp.119-139, 2006.

\bibitem
1E. Candes, J. Romberg, T. Tao,
\lq\lq Robust uncertainty principles: Exact signal reconstruction from highly incomplete frequency information,"
\emph{IEEE Transactions on Information Theory},
vol.52, pp.489-509, 2006.

\bibitem
1J. Bian, J. H. Siewerdsen, X. Han, E. Y. Sidky, J. L. Prince, C. A. Pelizzari, X. Pan,
\lq\lq Evaluation of sparse-view reconstruction from flat-panel-detector cone-beam CT,"
\emph{Physics in Medicine and Biology},
vol.55, pp.6575, 2010.

\end{thebibliography}
\end{document}